# COMPARING DECISION SUPPORT TOOLS FOR CARGO SCREENING PROCESSES

**Peer-Olaf Siebers[a], Galina Sherman[b], Uwe Aickelin[c], David Menachof[d]**

[a, c] School of Computer Science, Nottingham University, Nottingham NG8 1BB, UK.
[b, d] Business School, Hull University, Hull HU6 7RX, UK.

[a]pos@cs.nott.ac.uk, [b]g.sherman@2008.hull.ac.uk, [c]uxa@cs.nott.ac.uk, [c]d.menachof@hull.ac.uk

**ABSTRACT**

When planning to change operations at ports there are two key stake holders with very different interests involved in the decision making processes. Port operators are attentive to their standards, a smooth service flow and economic viability while border agencies are concerned about national security. The time taken for security checks often interferes with the compliance to service standards that port operators would like to achieve.

Decision support tools as for example Cost-Benefit Analysis or Multi Criteria Analysis are useful helpers to better understand the impact of changes to a system. They allow investigating future scenarios and helping to find solutions that are acceptable for all parties involved in port operations.

In this paper we evaluate two different modelling methods, namely scenario analysis and discrete event simulation. These are useful for driving the decision support tools (i.e. they provide the inputs the decision support tools require). Our aims are, on the one hand, to guide the reader through the modelling processes and, on the other hand, to demonstrate what kind of decision support information one can obtain from the different modelling methods presented.

Keywords: port operation, service standards, cargo screening, scenario analysis, simulation, cost benefit analysis, multi criteria analysis

## 1. INTRODUCTION

Businesses are interested in the trade-off between the cost of risk mitigation and the expected losses of disruptions (Kleindorfer and Saad 2005). Airports and seaports face an additional complexity when conducting such risk analysis. In these cases there are two key stake holders with different interests involved in the decision processes concerning the port operation or expansion (Bichou 2004).

On the one hand we have port operators which are service providers and as such interested in a smooth flow of port operations as they have to provide certain service standards (e.g. service times) and on the other hand we have the border agency which represent national security interests that need to be considered. Checks have to be conducted to detect threats such as weapons, smuggling and sometimes even stowaways. If the security checks take too long they can compromise the service standard targets to be achieved by the port operators. Besides these two conflicting interest there is also the cost factor for security that needs to be kept in mind. Security checks require expensive equipment and well trained staff. However, the consequences for the public of undetected threats passing the border can be severe. It is therefore in the interest of all involved parties to find the right balance between service, security, and costs.

But how can we decide the level of security required to guarantee a certain threshold of detection of threats while still being economically viable and not severely disrupting the process flow? A tool frequently used by business and government officials to support the decision making is Cost-Benefit Analysis (CBA) (Hanley and Spash 1993). While CBA is useful in our case to find the right balance between security and costs it struggles to provide decision support for the consideration of service quality. This is due to the fact that service quality is difficult to be expressed in monetary terms. Multi Criteria Analysis (MCA) is a tool that allows taking a mixture of monetary and non monetary inputs into account. It can use the results of a CBA as monetary input and service quality estimators as non monetary input and produce some tables and graphs to show the relation between cost/benefits of different options (DCLG 2009).

Different modelling methods can be used to conduct CBA. Damodaran (2007) lists Scenario Analysis (SA), Decision Trees (DT) and Monte Carlo Simulation (MCS) as being the most common ones in the field of Risk Analysis. Less frequently used in Risk Analysis but often used in Operational Research to investigate different operational practices is Discrete Event Simulation (DES) (Turner and Williams 2005; Wilson 2005). Depending on the world view the modeller adopts DES models can either be process oriented or object oriented (Burns and Morgeson 1988). Besides being useful for estimating factors required for CBA, DES models also allow to investigation how well service standards are reached in the system under investigation as it considers delays that one experiences while moving through the system (Laughery et al 1998). It is therefore well suited, in conjunction with CBA, to provide all the inputs required for a MCA. However, sometimes it is even possible to find a solution that does not require any investment. It might be feasible to achieve the goal simply by changing certain working routines. In such cases DES might be able to provide the information required for making a better informed decision without the need to conduct a full MCA.

In previous work (Sherman et al 2010) we compared the efficiency of different methods for conducting CBA by using the same case study. This strategy allowed us to contrast all modelling methods with a focus on the methods themselves, avoiding any distortions caused by differences in the chosen case studies. In this paper we continue our investigation but this time we focus on the two competing DES approaches and what information they can provide to assist our analysis.

The remainder of the paper is structured as follows. In Section 2 we introduce our case study system, the Port of Calais. In Section 3 we show in detail how to conduct a CBA using SA for our case study system. In Section 4 we discuss the additional features DES has to offer and we explain how to implement a model of the case study system using different world views. Once we discussed the specific features of the different implementations, we demonstrate by an experiment how to use DES to provide decision support. In Section 5 we summarise the findings from Sherman et al (2010) and this paper in form of a table that shows real world phenomena that can be modelled with the different modelling methods, the data requirements, and the decision support information that is provided. Finally we suggest some future research activities.

## 2. CASE STUDY

Our case study involves the cargo screening facilities of the ferry Port of Calais (France). In this area of the port there are two security zones, one operated by French authorities and one operated by the UK Border Agency (UKBA). The diagram in Figure 1 shows the process flow in this area.

According to the UKBA, between April 2007 and April 2008 about 900,000 lorries passed the border and approximately 0.4% of the lorries had additional human freight (UKBA 2008). These clandestines as they are called by the UKBA are people who are trying to enter the UK illegally - i.e. without having proper papers and documents.

The search for clandestines is organised in three major steps, one by France and two by the UKBA. On the French site all arriving lorries are screened, using passive millimetre wave scanners for soft sided lorries and heartbeat detectors for hard sided lorries. If lorries are classified as suspicious after the screening further investigations are undertaken. For soft sided lorries there is a second test with $CO_2$ probes and if the result is positive the respective lorry is opened. For hard sided lorries there is no second test and they are opened immediately.

Although 100% of lorries are screened at the French site, not all clandestines are found. This shows that the sensor efficiency in the field is less than 100%. Unfortunately it is not known for any of the sensors how much less their efficiency is and field tests cannot be undertaken as it would be unethical to deliberately lock someone in the back of a lorry. Another problem with estimating the sensor efficiency is that the sensor data has to be interpreted by human operators, who might misinterpret them. Again, no data exist about the efficiency of operators.

On the British site only a certain percentage of lorries (currently 33%) is searched at the British sheds. Here a mixture of measures is used for the inspection, e.g. $CO_2$ probes, dogs, and opening lorries. Once the lorries passed the British sheds they will park in the berth to wait for the ferry. In the berth there are mobile units operating that search as many of the parked lorries as possible before the ferry arrives, using the same mixture of measures than in the sheds. As shown in Table 1 only about 50% of the clandestines detected were found by the French, about 30% in the sheds and 20% by the mobile units in the berth. The overall number of clandestines that are not found by the authorities is of course unknown.

Table 1: Statistics from Calais

| Statistic | Value |
|---|---|
| Total number of lorries entering Calais harbour | 900,000 |
| Total number of positive lorries found | 3474 |
| Total number of positive lorries found on French site | 1,800 |
| Total number of positive lorries found on UK site | 1,674 |
| … In UK Sheds | 890 |
| … In UK Berth | 784 |

The key question is: What measures should we employ to reduce the overall number of clandestines that make it through to the UK? One way to improve detection rates could be to intensify the search activities. As we can see, clandestines are found at all stages of the cargo screening process and we can be sure that not all clandestines will be found in the end. However, when increasing search activities we also have to consider the disruptions this might cause to traffic flow. As mentioned in Section 1 we have different stakeholders with different interests involved in the decision making process. The two key objectives on which to base the decision are as follows: minimise costs (for the tax payer) and maximise service quality (by minimising waiting times and disruptions)

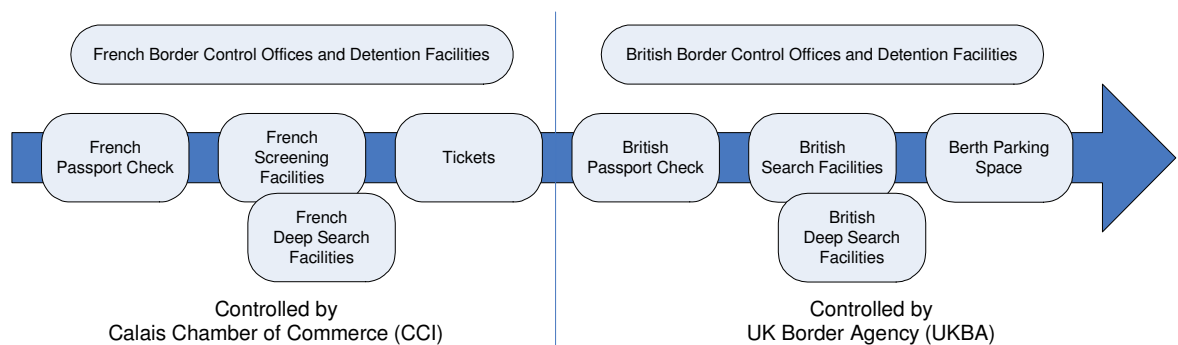

Figure 1: Border control operations at Calais

## 3. USING CBA & SA FOR DECISION SUPPORT

CBA seeks to value the expected impacts of an option in monetary terms (DCLG 2009). It involves comparing the Total Expected Costs (TEC) of each option against the total expected benefits, to see whether the benefits outweigh the costs, and by how much (Nas 1996). The aim is to determine the efficiency of the interventions relative to each other and the status quo. In our case total expected costs comprise the investments we have to make to increase the search activities. This might include things like increasing staff level, staff training, getting new sensors with better technology or building new search facilities. The total expected benefits will be the money that is saved for each clandestine that does not make it into the UK. Clandestines that made it to the UK are very likely to work illegally and therefore causing some income tax losses. Furthermore, they will not pay their contributions to health insurance and pensions. Therefore, the government will have to look after them once they are ill or old.

Uncertainty in the CBA parameters is often evaluated using a sensitivity analysis, which shows how the results are affected by changes in the parameters. In our case we have one parameter for which it is impossible to collect any data: the number of positive lorries that make it into the UK. A positive lorry is a lorry that has at least one clandestine on board. Usually clandestines attempt to cross the border in small groups. Conducting a sensitivity analysis cannot solve this problem but it can show the impact of changing the key parameter on the decision variable and can consequently provide some additional information to aid the decision making process and improve confidence in the decision made in the end.

SA is the process of analysing possible future events by considering alternative possible outcomes (Refsgaarda et al 2007). To define our scenarios we consider the following two factors: Traffic Growth (TG) and Positive Lorry Growth (PLG). For each of the factors we define three scenarios. Table 2 shows the factors and scenarios investigated and an estimate of the likelihood for each of the scenarios to occur in the future. A justification for the scenario definitions can be found in the following two paragraphs.

Table 2: Two factors with three scenarios each and their probability of occurrence

| Factor 1 | TG | p(TG) |
|---|---|---|
| Scenario 1 | 0% | 0.25 |
| Scenario 2 | 10% | 0.50 |
| Scenario 3 | 20% | 0.25 |
| Factor 2 | PLG | p(PLG) |
| Scenario 1 | -50% | 0.33 |
| Scenario 2 | 0% | 0.33 |
| Scenario 3 | 25% | 0.33 |

Our TG scenarios are based on estimates by the port authorities who are planning to build a new terminal in Calais in 2020 to cope with all the additional traffic expected. According to DHB (2008) between 2010 and 2020 the traffic in the Port of Dover is expected to double. We assume that this is also applicable to the Port of Calais and have therefore chosen the 10% per annum growth scenario as the most likely one, while the other two are equally likely.

Our PLG scenarios are reflecting possible political changes. The number of people trying to enter the UK illegally depends very much on the economical and political conditions in their home countries. This is difficult to predict. If the situation stabilises then we expect no changes in the number of attempts to illegally enter the UK. However, as our worst case scenario we assume a 25% growth. Another factor that needs to be considered is that trafficking attempts very much depend on the tolerance of the French government to let clandestines stay nearby the ferry port while waiting for the opportunity to get on one of the lorries. However, it is currently under discussion that the French authority might close the camps where clandestines stay which would reduce the number of potential illegal immigrants drastically. Therefore we added a scenario where the number of attempts is drastically reduced by 50%. As there is currently no indication of which of these scenarios is most likely to occur we have assigned them the equal probabilities of occurrence. We will assume that any changes in clandestine numbers will proportionally affect successful and unsuccessful clandestines.

The question that needs to be answered here is how the UKBA should respond to these scenarios. We assume that there are three possible responses: not changing the search activities, increasing the search activities by 10% or increasing the search activities by 20%. For the CBA Search Growth (SG) is our primary decision variable.

The cost for increasing the search activities in Calais is difficult to estimate, as there is a mixture of fixed and variable cost and operations are often jointly performed by French, British and private contractors. However, if we concentrate on UKBA's costs, we can arrive at some reasonable estimates, if we assume that any increase in searches would result in a percentage increase in staff and infrastructure cost. Thus we estimate that a 10% increase in search activity would cost £5M and a 20% increase £10M.

Now we need to estimate the benefits we would expect when increasing the search activities. First we need to derive a figure for the number of Positive Lorries Missed (PLM) and how much each of these lorries cost the tax payer. A best guess of "successful" clandestines is approximately 50 per month (600 per year). With an average of four clandestines on each positive lorry an estimated 150 positive lorries are missed each year. It is estimated that each clandestine reaching the UK costs the government approx. £20,000 per year. Moreover, UKBA estimates that the average duration of a stay of a clandestine in the UK is five years, so the total cost of each clandestine slipping through the search in Calais is £100,000, resulting in £400,000 per PLM.

It is probably a fair assumption that an increase in searches will yield a decrease in the number of positive lorries and an increase in traffic will yield an increase in PLM. In absence of further information we assume linear relationships between the two parameters. Table 3 shows the number of PLM) assuming there is no PLG. Equation 1 has been used to produce the table.

$$PLM(TG,SG)=PLM*(1+TG)/(1+SG) \quad (1)$$

Table 3: PLM for (PLG=0)

| PLG 0% | SG 0% | SG +10% | SG +20% |
|---|---|---|---|
| TG 0% | 150.00 | 136.36 | 125.00 |
| TG 10% | 165.00 | 150.00 | 137.50 |
| TG 20% | 180.00 | 163.64 | 150.00 |

With this information we can now calculate the Economic Cost (EC) for all the different scenarios (see Table 4) using Equation 2:

$$EC(TG,SG,PLG)=PLM(TG,SG)*(1+PLG) \quad (2)$$

Table 4: EC for different SG options

| SG 0% | PLG -50% | PLG 0% | PLG 25% |
|---|---|---|---|
| TG 0% | £30,000,000 | £60,000,000 | £75,000,000 |
| TG 10% | £33,000,000 | £66,000,000 | £82,500,000 |
| TG 20% | £36,000,000 | £72,000,000 | £90,000,000 |
| SG 10% | PLG -50% | PLG 0% | PLG 25% |
| TG 0% | £27,272,727 | £54,545,455 | £68,181,818 |
| TG 10% | £30,000,000 | £60,000,000 | £75,000,000 |
| TG 20% | £32,727,273 | £65,454,545 | £81,818,182 |
| SG 20% | PLG -50% | PLG 0% | PLG 25% |
| TG 0% | £25,000,000 | £50,000,000 | £62,500,000 |
| TG 10% | £27,500,000 | £55,000,000 | £68,750,000 |
| TG 20% | £30,000,000 | £60,000,000 | £75,000,000 |

To be able to calculate the benefit we need to know the combined probabilities of each scenario's likelihood to occur (listed in Table 2). We get this by multiplying the probabilities of the individual scenarios as shown in Equation 3. The results of these calculations can be found in Table 5.

$$p(TG,PLG)=p(TG)*p(PLG) \quad (3)$$

Table 5: Combined probabilities

|  | PLG -50% | PLG 0% | PLG 25% |
|---|---|---|---|
| TG 0% | 0.0833 | 0.0833 | 0.0833 |
| TG 10% | 0.1667 | 0.1667 | 0.1667 |
| TG 20% | 0.0833 | 0.0833 | 0.0833 |

Now we multiply the EC from Table 4 with the probabilities from Table 5 to receive the TEC for each SG option, using Equation 4. The results are shown in Table 6. The final step is to calculate the Net Benefit (NB) by using SG=0 as the base case. The NB can be calculated using Equation 5 (where C = cost for SG).

$$TEC(SG)=\sum(EC(SG,TG,PLG)*p(TG,PLG)) \quad (4)$$

$$NB(SG)=TEC(SG=0)-TEC(SG)-C(SG) \quad (5)$$

Table 6: CBA for different SG options

| Option | 1 | 2 | 3 |
|---|---|---|---|
| SG | 0% | 10% | 20% |
| TEC | £60,500,000 | £55,000,000 | £50,416,667 |
| C | £0 | £5,000,000 | £10,000,000 |
| NB | £0 | £500,000 | £83,333 |

The results of the CBA suggest that there is a small benefit in choosing option 2. However, we need to keep in mind that the calculations are based on a lot of assumptions. Therefore, a small difference in the NB might just be random noise. In particular we have one factor that we do not know anything about - PLM. In order to learn more about the impact of this factor we can conduct a sensitivity analysis. Running CBA for several different PLMs reveals that for a higher PLM option 3 would give us the most benefits but for a lower PLM option 1 would be the best choice. The sensitivity analysis shows how important it is to get a good estimate of the PLM.

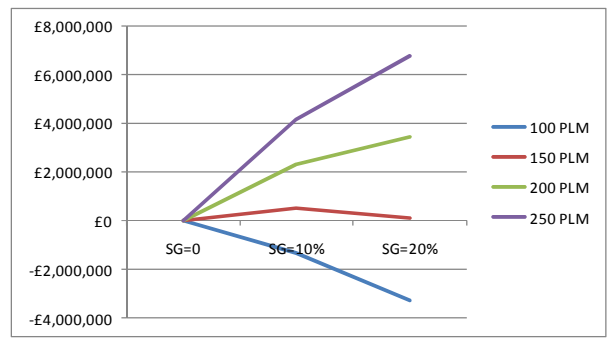

Figure 2: Sensitivity analysis results

If nothing else, the process of thinking through scenarios is a very useful exercise. In our case SA helped us to define the scenarios of interest. The sensitivity analysis later helped us to understand the importance of PLM when choosing an option. Overall SA only allows investigating a small number of factors but it seems to be a useful tool for structuring the overall investigation and get a first estimate regarding the benefits that one could gain from the different option.

## 4. USING DES FOR DECISION SUPPORT

The main benefit of DES models is that time and space can be taken into account which allows us for the first time to assess service quality (in terms of waiting time) and consider real world boundaries (e.g. space limitations for queues). As we said before our goal is to find a balance between service quality and security. CBA on its own or in conjunction with any of the methods introduced before does not provide us with information about the impact of changes on service quality. Another benefit we get by using DES modelling is that it simplifies adding more operational details about the real system and better supports the utilisation of real world data which both make the models themselves and results more credible.

In the following two sub sections we describe two different DES modelling approaches: Process Oriented DES (PO DES) and Object Oriented DES (OO DES). The differences between these methods will be described later in the subsequent sections. Here we look at commonalities. For both modelling approaches we need additional data regarding arrival rates, cycle times (time lorries spend in a shed for screening), space availability between stations for queuing, and resources for conducting the searches.

In order to be able to emulate the real arrival process of lorries in Calais we created hourly arrival rate distributions for every day of the week from a year worth of hourly number of arrival records that we received from UKBA. These distributions allow us to represent the real arrival process, including quiet and peak times. In cases where this level of detail is not relevant we use an exponential distribution for modelling the arrival process and the average arrival time calculated from the data we collected as a parameter for this distribution.

The cycle times are based on data that we collected through observations and from interviews with security staff. In order to represent the variability that occurs in the real system we use different triangular distribution for each sensor types. Triangular distributions are continuous distributions bounded on both sides. In absence of a large sample of empirical data a triangular distribution is commonly used as a first approximation for the real distribution (XJTEK 2005). Every time a lorry arrives at a shed a value is drawn from a distribution (depending on the sensor that will be used for screening) that determines the time the lorry will spend in the shed.

We did not put any queue size limits into our case study simulation model. However, we have a run time display variable for each individual queue that displays the maximum queue length of that queue. In this way we can observe which queue is over its limit without interrupting the overall simulation run. If necessary, queue length restrictions could easily be added. In fact, in one of our experiments we restrict the capacity of UK shed queue and let lorries pass without searching them if the queue in front of the shed is getting too long. This strategy improves the service quality but has a negative impact on security. Simulation allows us to see how big the impact of this strategy is in the different scenarios.

We also added some more details to the UK berth operation representation. We now consider the (hourly) arrival of the ferry boat. When the ferry arrives all search activities in the berth area are interrupted and all lorries currently in the berth area are allowed to board the ferry (as long as there is enough space on the ferry), regardless if they have been checked or not. Again, this strategy improves the service quality but has a negative impact on security. This is an optional feature of the DES model that can either be switched on or off.

Finally, in DES we can consider resources in a more detailed way. While previously resources have only been playing a role as a fixed cost factor (cost for SG), we can now take into account how different resource quantities influence the process flow and subsequently the service quality. These quantities can vary between different scenarios, but also between different times of day (e.g. peak and quiet times).

Clearly, as we can see from the above, DES helps to improve the decision making process. It allows besides the monetary estimates to get information on service quality and gain further insight into system operations. This additional insight can be used for decision making but also for optimising system operations. DES also allows you to conduct a sensitivity analysis to find high impact factors that need to be estimated more carefully. Of course this does not come without costs. DES models usually take much longer to build and need additional empirical data.

### 4.1. DES using a process oriented world view

As before, we use the standard procedure for calculating the TEC for the different SG options that enables us to conduct a basic CBA. For this the PO DES delivers the number of Positive Lorries Found (PLF) which allows us to calculate the number of PLM under the assumption that there is a linear relationship between these two measures. This number can then be used as an input for the CBA.

On a first view the PO DES model looks very similar to the MCS model presented in Sherman et al (2010) and in fact it is an extended version of this model. In addition to the routing information defined in the MCS model we have added arrival rates, cycle times for screening the lorries and available resources. The data required for implementing these additions into the model has been provided by UKBA.

We now have a stochastic and dynamic simulation model that allows us to observe the evolution of the system over time. This is a big benefit compared to the static models we used previously as it allows us for the first time to consider our second objective in the analysis - to provide high quality service. One of the key factors for providing high quality service is to keep average waiting times below a specified threshold (service standard). By adding arrival rates, cycle times and available resources the model is able to produce realistic waiting time distributions, which gives us an indication of the service quality we are achieving with different parameter settings.

Another useful output that PO DES provides is resource utilisation. This information allows us to fine-tune our CBA as we are able to better estimate SG costs. So far we have assumed that SG has a fixed cost which linear correlated with TG. In reality however, the cost for SG might depend on the utilisation of the existing resources. If in the current situation facility and staff utilisation is low then SG can perhaps be implemented without any additional costs. PO DES allows to test at what level of TG additional resources are required (throughput analysis).

The PO DES model also allows us to analyse queue dynamics. A useful statistic to collect in this context is the "maximum queue sizes" which enables us to find bottlenecks in our system. With this information we can then pinpoint where in the system we should add additional resources to achieve maximum impact. Removing bottlenecks improves the system flow and consequently service quality.

Another interesting feature of our simulation model is that it allows us to represent temporal limited interventions and see what their effect is on system flow and detection rates of positive lorries. These procedures could be officers speeding up when queues are getting longer (missing more positive lorries) or changing search rates over time (less search at peak times) or stopping search completely once a certain queue length has been reached. PO DES can help us to find out, which of these strategies is best.

However, the process oriented modelling approach as described above has some limitations with regards to flexibility. We are using proportions for defining the vehicle routing (and detection rates), based on historic data. This assumes that even if we have changes in the system these proportions would always remain the same. While this is acceptable for many situations (in particular for smaller changes) there are occasions where this assumption does not hold. For example, if we change the proportion of vehicles searched in the berth from 5% to 50% we cannot simply assume that the detection rate grows proportionally. While it might be easy to spot some positive lorries as they might have some signs of tampering (so the detection rate for these is very high and these where the ones reported in the historic data), it will get more difficult once these have been searched and a growth in search rate does not yield an equivalent growth in detection rate any more. This needs to be kept in mind when using a PO DES approach.

Furthermore, the assumption of a linear relationship between number of PLF and number of PLM which is one of our key assumptions for the CBA is quite weak in connection with PO DES as this relationship can be influenced by some of the interventions. For example, the temporal limited intervention of stopping search completely once a certain queue length has been reached influences the overall number of clandestines in the system (when calculating this value by adding up the number of PLF and number of PLM), although this number should not be influenced by any strategic interventions. This needs to be considered in the analysis of the results..

### 4.2. DES using an object oriented world view

OO DES has a very different world view compare to PO DES. Here we model the system by focusing on the representation of the individual objects in the system (e.g. lorries, clandestines, equipment, or staff) rather than using predefined proportions for the routing of entities based on historic data. In fact we only use historic data as inputs (arrival rates, search rates, staffing) and for validation purposes (numbers of PLF at the different stages). The system definition is now based on layout information and assumptions on sensor capabilities. Taking this new perspective means that we transfer all the "intelligence" from the process definition into the object definition and therefore change our modelling perspective from top down to bottom up.

Unlike in DT, MCS, and PO DES we do not use a probabilistic framework for directing lorries (and deciding which of the lorries are positive). Instead we aim to recreate processes as they appear in the real system. At the entrance of the port we mark a number of lorry entities as positive (to simulate clandestines getting on board to lorries). This gives us complete control over the number of positive lorries entering the system. As the lorries go through the system they will be checked at the different stages (French side, UK sheds, UK berth). For these checks we use sensors that have a specific probability of detecting true and false positives. Only lorries that have been marked positive earlier can be detected by the sensors as true positives. The marked lorries that are not detected at the end (port exit) are the ones that will make it through to the UK. Officers can decide how to use the detectors (e.g. speed up search time if queues are getting to long, change search rates, etc.) depending on environmental conditions.

One of the advantages of this simulation method is that it is much easier to manipulate object and system parameters, like for example the sensor detection rates and search rates. When we change these parameters we do not have to worry about linear dependencies any more. In fact, we can find out relationships between variables by looking at the evolution of the system over time. However, the most interesting thing here is that due to the fact that we model sensors as objects with each having individual detection rates the number of PLM becomes an output of the simulation. Yet, we have to keep in mind that this output is still a function of unknowns: PLM = f(lorries marked positive, sensor detection rates ...). But overall, this is very useful information as we can now do some what-if analysis and see the trends of how changes in the system setup impact on number of PLM. We do not rely on the implicit assumption of a linear relationship between

PLM and SG any more. While this is not directly solving our problem of estimating how many positive lorries we miss it gives us some additional information about system behaviour that might help us to decide in one way or another.

### 4.3. Experimentation with the OO DES model

In this sub section we want to show how we can use our DES model to test alternative scenarios. We have implemented our model in AnyLogic[TM] 6.6, a Java[TM] based multi-paradigm simulation software. Figure 3 shows a screenshot of a section of the simulation model (berth area) during execution.

To set up our OO DES we tried to reproduce the base scenario (as defined in Table 1) as closely as possible by calibrating our model to produce the correct values for the number of PLF at the different stages of the search process. We can do this by varying the number of positive lorries entering the port, the sensor detection rates, and the berth search rate. The results of the calibration exercise are presented in Table 7 (Scenario 1). To get somewhere close to the real PLF values at the different stages we had to increase the number of positive lorries entering the port. Hence, also the PLM value is much higher than the best guess we used in our CBA. The detection rates for the UK sheds and the UK berth had to be much higher than the ones on the French side, in order to match the true rates. We assume that in the real system this is due to the fact that UKBA uses some intelligence when choosing the lorries to be screened. Therefore their success rate is much higher. In particular in the berth officers can drive around and pick suspicious lorries without any time pressure.

Our scenarios are defined by TG and SG. All other parameters (grey) depend on these values (with the exception of the queue size restriction). All dependencies are explained in Section 3. For this experiment we assume that there is no change in the number of people trying to get into the UK (PLG=0). In Table 7 we only show the changes in the scenario setup. Empty fields mean that there is no change in the set-up for that specific parameter.

For this experiment we have defined a service standard that needs to be achieved. We have a threshold time in system that should not be exceeded by more than 5% of the lorries that go through the system. We have one intervention that allows us to influence the process flow. We can define a threshold for the maximum queue size in front of the UK sheds (queue size restriction). If this threshold is exceeded lorries are let pass without screening. While this intervention improves the flow there is a risk that more positive lorries are missed as less lorries are inspected.

The first three scenarios deal with TG. There is no problem in regards to compliance with service standards. Resource utilisation does not change as the number of searches does not change. However, the number of PLF is decreasing while the number of PLM is going up. The next two scenarios deal with SG. The increase in search activities causes some delays and at a SG of 20% the system does not comply with service standards any more. On the other hand, as expected, an increase in search activities improves the number of PLF and reduces the number of PLM. Scenario 5 indicates that service standards cannot be achieved with the current staff/facilities and that an investment has to be made in order to reduce the number of PLM and comply with service standards. However, the last scenario shows that there is also a strategic solution possible that does not require any investment. By managing the queues in front of the UK sheds it is possible to reduce the number of service problems to fewer than 5% (compliant with service standards) while still keeping the number of PLM at a very low level. Therefore, when applying this intervention no investment is required.

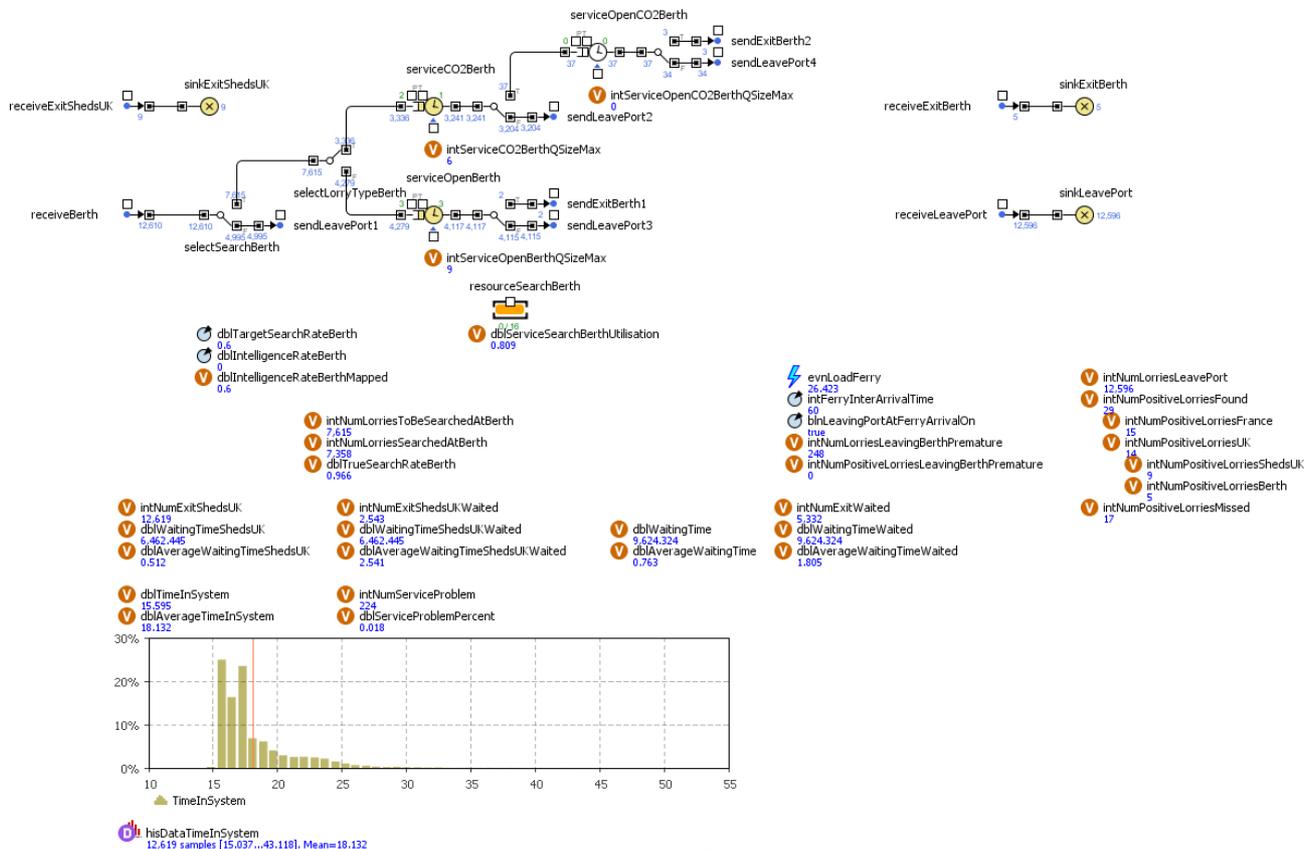

Figure 3: OO DES running in AnyLogic[TM] (screenshot of berth area only)

Table 7: OO DES simulation experiment set-ups and results (10 replications)

| Scenarios | | 1 | 2 | 3 | 4 | 5 | 6 | 7 |
|---|---|---|---|---|---|---|---|---|
| Traffic Growth (TG) | | 0% | 10% | 20% | 0% | | | |
| Search Growth (SG) | | 0% | | | 10% | 20% | | |
| Lorries | Arrivals | 900000 | 990000 | 1080000 | 900000 | | | |
| | Soft-sided | 0.44 | | | | | | |
| | Positive | 0.00550 | 0.00500 | 0.00458 | 0.00550 | | | |
| Search rate | UK Sheds | 0.330 | 0.300 | 0.275 | 0.363 | 0.396 | | |
| | UK Berth | 0.600 | 0.545 | 0.500 | 0.660 | 0.720 | | |
| Detection Rates | France | 0.41 | | | | | | |
| | UK Sheds | 0.80 | | | | | | |
| | UK Berth | 0.95 | | | | | | |
| Queue size restriction | UK Sheds | off | | | | | 10 | 9 |
| **Results** | | 1 | 2 | 3 | 4 | 5 | 6 | 7 |
| Waiting times (avg)[*1)] | France | 0.858 | 1.019 | 1.268 | 0.863 | 0.859 | 0.860 | 0.863 |
| | UK Sheds | 2.612 | 2.474 | 2.321 | 3.452 | 5.046 | 3.940 | 3.763 |
| | Overall | 1.831 | 1.783 | 1.856 | 2.439 | 3.620 | 2.901 | 2.788 |
| Time in system (avg) | | 18.099 | 18.085 | 18.155 | 18.517 | 19.274 | 18.893 | 18.834 |
| Service problem | | 0.019 | 0.019 | 0.020 | 0.036 | 0.068 | 0.052 | 0.049 |
| Resource utilisation | UK Sheds | 0.676 | 0.676 | 0.677 | 0.744 | 0.812 | 0.803 | 0.801 |
| | UK Berth | 0.808 | 0.808 | 0.809 | 0.868 | 0.915 | 0.914 | 0.914 |
| Positive lorries | France | 1774.9 | 1765.5 | 1745.9 | 1780.5 | 1774.3 | 1757.5 | 1769.7 |
| | UK Sheds | 900.8 | 814.0 | 733.8 | 981.2 | 1078.0 | 1061.2 | 1042.8 |
| | UK Berth | 699.9 | 658.4 | 630.7 | 715.9 | 743.0 | 746.5 | 746.8 |
| | Missed | 1590.1 | 1697.2 | 1797.0 | 1480.7 | 1365.7 | 1361.7 | 1358.1 |

As we have found a solution for our problem that does not require balancing costs and benefits there is no need to conduct a MCA. However, for other scenarios where we are not so lucky we might want to consider using MCA. A good guide to MCA with a worked example is provided by DCLG (2009).

## 5. CONCLUSIONS

To help the managers and analysts to decide which method best to use for supporting their decision processes concerning port operation and expansion it is important to be clear about the real world phenomena that can be considered with the different methods and the decision support information that can be obtained by applying them. Table 8 list these for all the methods discussed in Sherman et al (2010) and in this paper.

The methods can be divided in two categories: static and dynamic. For static methods the passage of time is not important (Rubinstein and Kroese 2008). SA, DT, and MCS belong to this category. The lack of a concept of time in these methods makes is impossible to analyse service quality of a system as all performance measures for such analysis rely on the passage of time. On the contrary, dynamic methods consider the passage of time. DES belongs to this category and allows evaluate a system's compliance with service standards. Besides, it provides other useful information about the dynamics of the system that can be used for optimising processes and improving flows.

DES can be implemented in different ways, either as PO DES or OO DES. In our experience PO DES seems to be easier to implement but less flexible (easier to manipulate). OO DES seem to be more flexible but also more difficult to implement. However, it has some advantages. In our case OO DES was the only tool that also provided an estimate of the number of PLM, which helps us to better evaluate the effect of different interventions.

For evaluating the trade-off between security and cost CBA is a valuable tool. If we want to balance security, cost and service then MCA is a better choice. While CBA relies on monetary inputs, MCA allows using monetary and non-monetary inputs. This is useful as service quality is difficult to capture by monetary measures. However, sometimes none of these tools is required for the analysis as the answer might come directly from the model as we have shown in Section 4.3.

A natural extension of the OO DES modelling approach would be to add "intelligent" objects that have a memory, can adapt, learn, move around, and respond to specific situations. These could be used to model officers that dynamically adapt their search strategies based on their experiences, but also clandestines learning from failed attempts and improving their strategies when trying again. We are currently working on implementing such "intelligent" objects.

Table 8: Comparison of modelling methods regarding real world phenomena that can be considered in the model (black) and decision support information that can be obtained from the model (red)

| SA | DT | MCS | PO DES | OO DES |
|---|---|---|---|---|
| Scenarios (factors and the decision variable) | Scenarios (factors and the decision variable) | Scenarios (factors and the decision variable) | Scenarios (factors and the decision variable) | Scenarios (factors and the decision variable) |
| Linear relationships | Linear relationships | Linear relationships | Linear relationships | - |
| TEC | TEC | TEC | TEC | TEC |
| PLM, PLF | PLM, PLF | PLM, PLF | PLM, PLF | PLM, PLF |
|  | System structure | System layout | System layout | System layout |
|  | Existing resources | Existing resources | Existing resources | Existing resources |
|  |  | System variability | System variability | System variability |
|  |  |  | Service time distributions | Servic time distributions |
|  |  |  | Resource utilisation | Resource utilisation |
|  |  |  | Dynamic system constraints (e.g. peak times) | Dynamic system constraints (e.g. peak times) |
|  |  |  | System throughput | System throughput |
|  |  |  | Waiting times (service quality) | Waiting times (service quality) |
|  |  |  | Time in system | Time in system |
|  |  |  | Bottleneck analysis | Bottleneck analysis |
|  |  |  | Dynamic decisions by system | Dynamic decisions by objects |
|  |  |  |  | Sensor detection rates |
|  |  |  |  | Number of pos. lorries entering the system |